\pdfoutput=1
\documentclass{article}

\usepackage{arxiv}

\usepackage[utf8]{inputenc}
\usepackage[T1]{fontenc}
\usepackage{natbib}
\usepackage{hyperref}
\usepackage{url}
\usepackage{amsmath,amssymb,bm}
\usepackage{booktabs}
\usepackage{graphicx}
\usepackage{multirow}
\usepackage{xspace}
\usepackage{microtype}

\usepackage{amsmath,amsfonts,bm}

\def\eqref#1{equation~\ref{#1}}

\def\1{\bm{1}}

\DeclareMathAlphabet{\mathsfit}{\encodingdefault}{\sfdefault}{m}{sl}
\SetMathAlphabet{\mathsfit}{bold}{\encodingdefault}{\sfdefault}{bx}{n}

\newcommand{\NUM}[1]{#1}

\newcommand{\INO}{INO\xspace}

\title{Interpolating Neural Operator (INO): A Data-Free and Efficient Approach for Learning PDE Solution Operators}

\renewcommand{\shorttitle}{Interpolating Neural Operator (INO)}

\author{
  Jiachen Guo \\
  Department of Mechanical Engineering \\
  Northwestern University \\
  HIDENN-AI, INC \\
  Evanston, IL 60201, USA \\
  \texttt{jiachen.guo@northwestern.edu} \\
  \texttt{jaguo@hidenn-ai.com} \\
  \And
  Ye Lu \\
  Department of Mechanical Engineering \\
  University of Maryland, Baltimore County \\
  Baltimore, MD 21250, USA \\
  \texttt{yelu@umbc.edu} \\
  \AND
  Naichen Shi \\
  Department of Industrial Engineering \\
  \& Management Sciences \\\& Mechanical Engineering \\
  Northwestern University \\
  Evanston, IL 60201, USA \\
  \texttt{naichen.shi@northwestern.edu} \\
  \And
  Thomas J.R. Hughes \\
  Oden Institute for Computational \\Engineering and Sciences \\
  University of Texas at Austin \\
  Austin, TX 78712, USA \\
  \texttt{tjr\_hughes@hotmail.com} \\
  \AND
  Wing Kam Liu \\
  Department of Mechanical Engineering \\
  Northwestern University \\
  HIDENN-AI, INC \\
  Evanston, IL 60201, USA \\
  \texttt{w-liu@northwestern.edu}
}

\begin{document}

\maketitle

\begin{abstract}
Neural operators have become a popular approach to approximate the solution operators of
parametric partial differential equations (PDEs). However, existing neural operators either require
a large amount of simulation data or a long physics-informed training on GPUs, and they cannot tell
how accurate an individual prediction is. In this paper, we propose the
Interpolating Neural Operator (INO), a data-free interpolating neural network that is trained
directly on the weak form of the PDE. In INO, the Karhunen--Lo\`eve coordinates of the input field
are treated as additional inputs together with the spatial coordinates, and each input is
approximated by a C-HiDeNN sub-network whose trainable parameters are nodal values. Since the
network is multilinear in its parameters, training reduces to a sequence of one-dimensional linear
solves by greedy alternating least squares. As a result, INO trains on one CPU core and predicts a new solution in microseconds. For coercive problems, the total error of every prediction is bounded by a computable
residual bound that requires no reference solution, and the same bound applies to the predictions of
other methods that satisfy the boundary conditions exactly. Before each prediction, INO checks whether the leading coordinates of the input lie within the range on which it is trained, and inputs outside this range can be passed to a conventional solver or to an INO trained on a wider range. INO is compared with physics-informed FNO and DeepONet on
different benchmarks. INO is the most accurate model on most of
these problems, by \NUM{15}$\times$ on two-dimensional Helmholtz at $65^2$ and \NUM{53}$\times$ on the
diffusion--reaction benchmark, and on the one- and two-dimensional problems its training on one
CPU core takes \NUM{3}--\NUM{80}$\times$ less time than the physics-informed baselines on one GPU.
\end{abstract}

\section{Introduction}

Many engineering tasks, such as design optimization, inverse identification and uncertainty
quantification, require solving a PDE for many different input functions. Conventional numerical solvers such as the finite element method have to run a new simulation for every input, which becomes
expensive when thousands of evaluations are needed. Data-driven neural operators such as DeepONet
and the Fourier neural operator (FNO) \citep{lu2021learning,li2020fourier} learn the solution
operator from pairs of input and solution fields. However, every training sample requires a full
numerical simulation. Physics-informed variants \citep{raissi2019physics,wang2021learning,
li2021physics} remove the need for data by minimizing the PDE residual, but they typically require long training time for fine resolution and 3D problems with stochastic optimization, and their errors
vary across random initializations (Appendix~\ref{app:variability}). None of these methods can tell
how accurate a particular prediction is.

\begin{figure}[t]
\begin{center}
\includegraphics[width=0.99\linewidth]{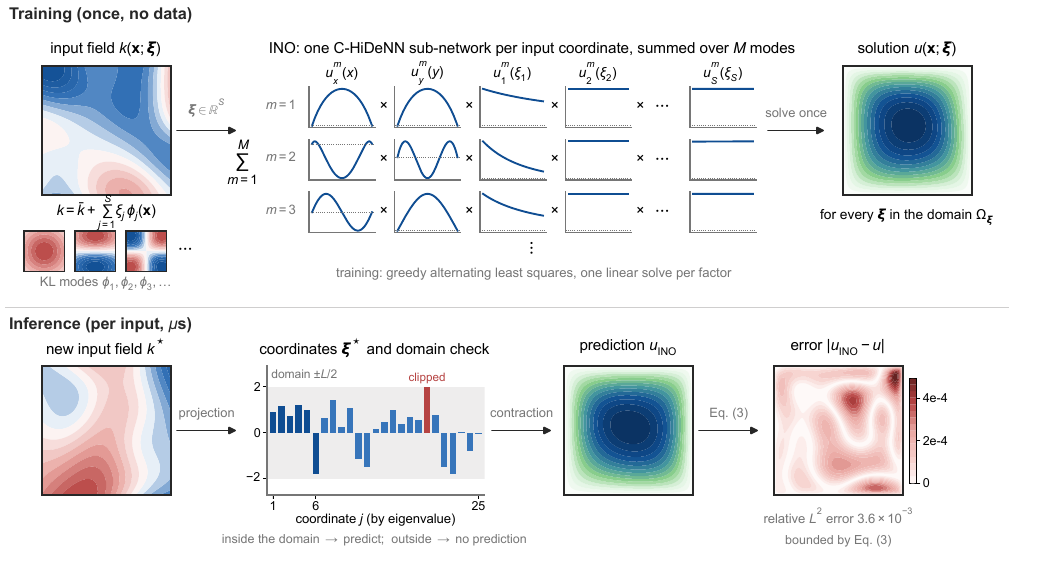}
\end{center}
\caption{Overview of INO, shown on 2D diffusion problem. \textbf{Training}: the input field is
represented by its $S$ KL coordinates $\bm\xi$, each coordinate becomes an input of the network with
its own C-HiDeNN sub-network (the curves are the one-dimensional factors of three modes of the
trained operator of Table~\ref{tab:neural}), and the weak form is solved once over the parametric
domain without training data. \textbf{Inference}: a new input field $k^\star$ is projected onto its
coordinates $\bm\xi^\star$; if its leading coordinates lie within the parametric domain (grey band), the prediction is
obtained by contracting the factors, and its error is bounded by Eq.~(\ref{eq:cert}) evaluated for
the true input field $k^\star$.}
\label{fig:overview}
\end{figure}

In this paper, we propose the Interpolating Neural Operator (INO), a data-free interpolating
neural network for the solution operators of parametric PDEs. As shown in Fig.~\ref{fig:overview}, the input field
is first represented by the coordinates of a truncated Karhunen--Lo\`eve (KL) expansion. These
coordinates are treated as additional inputs of the network together with the spatial coordinates,
so that a single network over space and parameters represents the solution operator. Each input is
fed into a convolution hierarchical deep-learning neural network (C-HiDeNN) sub-network
\citep{lu2023convolution,park2024engineering}, whose connectivity and activation functions are fixed
by a one-dimensional mesh and whose trainable parameters are nodal values. The sub-networks of all
inputs are multiplied to form a rank-one term, called a mode, and $M$ modes are summed. The
network is multilinear in its parameters, so
each training step is a linear solve of one-dimensional size, and training reduces to greedy
alternating least squares, which requires no data and no stochastic optimizer. The separated form is compact, and inference is a contraction of
one-dimensional factors (Section~\ref{sec:efficiency}). Since the C-HiDeNN interpolant has the Kronecker delta property, INO satisfies the boundary conditions exactly,
and its predictions can be substituted back into the PDE to bound the total error.

The main contributions of this paper are summarized as follows:

\begin{itemize}\itemsep1pt
    \item An efficient data-free operator, demonstrated on a moving heat source, diffusion and
        Helmholtz problems up to $513^2$ and $129^3$, a nonlinear diffusion problem and a space--time
        diffusion--reaction benchmark. On the one- and two-dimensional problems, its training by
        linear solves takes less time than physics-informed training, and its error depends little on
        the random initialization (Sections~\ref{sec:neural} and~\ref{sec:efficiency}).
    
    \item A computable error bound for every prediction (Section~\ref{sec:bound-theory}). The bound
        uses the true input field rather than its truncated expansion, so it covers the total
        error, needs no reference solution, and holds outside the parametric domain. It applies to
        any prediction that satisfies the boundary conditions, including those of a
        physics-informed FNO (Section~\ref{sec:accuracy}).
    
    \item A domain check that decides from the input alone whether INO makes a prediction, so that
        inputs outside the parametric domain can be passed to a conventional solver or to an INO
        trained on a wider parametric domain (Sections~\ref{sec:query} and~\ref{sec:neural}).
    
  \item A controlled comparison with physics-informed FNO and DeepONet on identical test inputs against
        reference solutions for the true input field, with each baseline trained with its original
        settings (Section~\ref{sec:neural}).
  \item An analysis of the error of INO, which shows that its spatial modes are accurate and that
        the error comes from how they are weighted for a given input; computing these weights by
        a small linear solve reduces the error by one to two orders of magnitude on the linear
        problems (Section~\ref{sec:efficiency}).
\end{itemize}

\section{Related work}
\label{sec:related}

Operator-learning methods such as DeepONet \citep{lu2021learning}, FNO \citep{li2020fourier} and
grid- or graph-based encoders \citep{gao2021phygeonet,pfaff2020learning,huang2023introduction}
learn the solution operator from pairs of input and solution fields generated by a conventional
numerical solver. Physics-informed variants replace the data by the PDE residual, which is evaluated at
collocation points in the physics-informed DeepONet (PI-DeepONet)
\citep{raissi2019physics,wang2021learning} or on the output grid in the physics-informed FNO
(PI-FNO) \citep{li2021physics}. These methods are data-free like INO, but their training is
a nonconvex optimization with known failure modes \citep{wang2022and,krishnapriyan2021characterizing,
grossmann2024can,mcgreivy2024weak}. Separable PINNs \citep{cho2023separable} also factorize the
network by dimension, but they still rely on a collocation loss and a stochastic optimizer.

The canonical polyadic (CP) decomposition
approximates a multivariate function by a sum of products of univariate functions. It is commonly
computed by alternating least squares \citep{carroll1970analysis,kolda2009tensor} or by greedy
rank-one enrichment \citep{zhang2001rank}, and low-rank tensor methods have been widely used to
solve PDEs \citep{bachmayr2023low}. In computational mechanics, proper generalized decomposition
(PGD) treats parameters as extra coordinates \citep{ammar2006new,chinesta2011short,nouy2010priori},
including stochastic PDEs with KL inputs \citep{nouy2007generalized}, and interpolating neural
networks combine this idea with C-HiDeNN interpolation
\citep{lu2023convolution,li2023convolution,guo2024convolutional,guo2025interpolating,guo2026large,park2024engineering}. INO builds on the separated form and the greedy training of these
methods and on the C-HiDeNN factors of interpolating neural networks. Relative to these works, INO
makes predictions for arbitrary input fields, which are projected onto their KL coordinates and
checked against the parametric domain before any prediction, and it bounds the error of each
prediction with the true input field, as discussed below. We also compare INO with
physics-informed neural operators on meshes up to $129^3$.

Stochastic collocation \citep{bungartz2004sparse} and stochastic Galerkin methods with polynomial
chaos \citep{ghanem1991stochastic,cohen2010convergence} are the classical methods for parametric
PDEs. Collocation solves the PDE at a set of parameter points and interpolates between them, one
deterministic solve per point, and stochastic Galerkin methods solve
one coupled system for all polynomial coefficients. Residual-based a posteriori error bounds have
been developed for reduced-basis methods \citep{rozza2008reduced,binev2011convergence}, for PGD
\citep{ladeveze2011verification} and for stochastic Galerkin methods \citep{eigel2014adaptive}. The
bound of Section~\ref{sec:bound-theory} is of this type, but its residual is computed with the true
input field instead of its truncated expansion. As a result, it bounds the total error of a
prediction, including the error from truncating the input field, and remains valid outside the
parametric domain. Since it needs only the prediction and the input field, it also applies to other
models whose predictions satisfy the boundary conditions, which we use to bound the error of PI-FNO
(Section~\ref{sec:accuracy}).

\section{INO formulation}
\label{sec:method}

\subsection{Network architecture}
\label{sec:kl}
\label{sec:promote}

The input field is first represented by the truncated KL expansion of a given covariance, which
can be written as:
\begin{equation}
  k(\bm x;\bm\xi)\;=\;\bar k+\sum_{j=1}^{S}\phi_j(\bm x)\,\xi_j ,
  \qquad \xi_j\in[-L/2,\,L/2],
  \label{eq:kl}
\end{equation}
where $\bar k$ is the mean of the field; $S$ is the number of retained KL modes; $\phi_j$ is the
$j$-th KL mode scaled by the square root of its eigenvalue; $\xi_j$ is the $j$-th KL coordinate \citep{ghanem1991stochastic}; and $L$ is the
width of the parametric domain in each direction, measured in standard deviations of the
coordinates. As a result, a truncated input field is represented by a point $\bm\xi$ in the
parametric domain $\Omega_{\bm\xi}=[-L/2,L/2]^S$. In INO, the
$S$ coordinates are treated as inputs together with the $d$ spatial coordinates, so that the
solution $u(\bm x,\bm\xi)$ becomes a function of $d+S$ variables. Since a grid in $d+S$ dimensions is
not feasible, INO approximates the solution in separated form (Appendix~\ref{app:structure}), which can be written as:
\begin{equation}
  u(\bm x,\bm\xi)\;\approx\;\sum_{m=1}^{M}\;\prod_{i=1}^{d}u^{m}_{x_i}(x_i)\;
  \prod_{j=1}^{S}u^{m}_{j}(\xi_j),
  \label{eq:ansatz}
\end{equation}
where $M$ is the total number of modes; $u^m_{x_i}$ and $u^m_j$ are the univariate functions of the
$m$-th mode in the spatial direction $x_i$ and in the parametric direction $\xi_j$; and the amplitude
$c_m$ of each mode (Appendix~\ref{app:algebra}) is absorbed into one of its factors. Each univariate
function $u^{m}_{j}(\xi_j)$ is a C-HiDeNN sub-network \citep{guo2025interpolating} of the form
$\widetilde{\bm N}(x;s,a,p)\,\bm u$, where $\bm u$ contains the nodal values of a one-dimensional
mesh and $\widetilde{\bm N}$ is the row vector of C-HiDeNN basis functions, which is controlled by
the patch size $s$, the dilation parameter $a$ and the reproducing order $p$ \citep{lu2023convolution}. The sub-network is a
partially connected neural network whose hidden layers are the linear finite element basis functions
of the mesh and a convolution patch of radial and polynomial neurons, and whose only trainable
parameters are the nodal values (Appendix~\ref{app:chidenn}). With $n$ nodes per spatial direction and $n_\xi$ nodes per parameter
direction, INO has $M(d\,n+S\,n_\xi)$ parameters, a number that grows linearly with $S$, whereas a grid would require $n^{d}n_\xi^{S}$ unknowns. Since the basis functions satisfy the Kronecker delta property, Dirichlet boundary conditions are imposed exactly on the nodal values. Therefore, INO needs no boundary penalty term, whose weight has to be tuned in physics-informed methods \citep{wang2022and}. All integrals are computed using Gaussian quadrature.

This paper distinguishes three errors. The \emph{rank error} is measured against a
reference solution of the same discretization for the truncated field of Eq.~(\ref{eq:kl}). The
\emph{representation error} is the difference between this truncated field and the true input field.
The \emph{total error} is measured against a reference solution for the true input field itself,
and it is the error reported against the baselines.

\subsection{Data-free training}
\label{sec:boosting}

INO is trained on the weak form of the PDE integrated over the parametric domain: the nodal values
of Eq.~(\ref{eq:ansatz}) are chosen such that
$\int_{\Omega_{\bm\xi}}\mathcal A_k(v,u)\,\mathrm d\bm\xi=\int_{\Omega_{\bm\xi}}F(v)\,\mathrm d\bm\xi$
for every test function $v$ of the same form, where $\mathcal A_k(\cdot,\cdot)$ is the bilinear form
of the PDE for the input field $k=k(\cdot\,;\bm\xi)$ and $F(v)=\int_\Omega fv$ is its linear
form. Two properties make the training efficient. First, since $k$ is
affine in $\bm\xi$, the weak form of $-\nabla\!\cdot(k\nabla u)=f$ splits into exactly $d(S{+}1)$
separated terms, each of which is a product of one-dimensional matrices, so that no integral over
the $(d+S)$-dimensional domain is needed (Appendix~\ref{app:algebra}). Second, Eq.~(\ref{eq:ansatz})
is multilinear in its parameters. If all factors except one are fixed, the weak form is linear in the
nodal values of the remaining factor, which are therefore obtained from one linear system whose
size equals the number of nodes of that factor's mesh. Therefore, training alternates
over the $d+S$ input directions with one linear solve per step, which is the alternating least
squares algorithm for CP decompositions \citep{carroll1970analysis,kolda2009tensor}. Modes are added one at a time (greedy enrichment) \citep{zhang2001rank}: the $M$-th mode is the
rank-one sub-network that solves the weak form for the residual left by the previous modes
(Appendix~\ref{app:algebra}), as in
the boosting training of interpolating neural networks \citep{park2024engineering, guo2025interpolating} and in PGD
\citep{ammar2006new,nouy2010priori}, and the amplitudes of all modes are updated afterward. 

\subsection{Inference}
\label{sec:query}

Given a new input field $k^\star$, INO first projects it onto the KL modes to obtain its
coordinates $\bm\xi^\star$, and then evaluates the univariate functions at $\bm\xi^\star$ and
contracts them (Section~\ref{sec:efficiency}). If the leading coordinates of the projected input
fall outside the parametric domain, INO does not return a prediction by default; the user can then run a conventional numerical solver, evaluate INO
anyway and accept its prediction if the error bound of Section~\ref{sec:bound-theory} is small
enough, or train a new INO on a wider parametric domain. INO assumes the input fields are smooth, so they can be accurately represented by a truncated KL expansion. It also assumes that, when the input is a coefficient, it remains positive throughout the entire parametric domain to ensure the problem stays elliptic; no comparable requirement is needed when the input appears in the source term.

\subsection{Error bound}
\label{sec:bound-theory}

INO is constructed using the locally supported basis functions based on C-HiDeNN interpolation, so its prediction is an explicit mesh-based function. This makes it interpretable and allows it to be inserted back into the discrete PDE. Let $A(k)$ be the matrix of the bilinear form $\mathcal A_k$ on the finite element space and
$F$ the load vector, so that the discrete solution $u(k)$ for the true input field $k$ satisfies
$A(k)\,u(k)=F$, and let $K_0$ be any symmetric positive definite matrix chosen to measure the
error, with the norm $\|v\|_{K_0}=(v^\top K_0v)^{1/2}$ and its dual norm
$\|r\|_{K_0^{-1}}=(r^\top K_0^{-1}r)^{1/2}$. Here $u(k)$ and the prediction $u_M$ are identified
with their vectors of nodal values. If $v^\top A(k)\,v\ge\alpha(k)\,v^\top K_0\,v$ holds for
every nodal vector $v$ with a coercivity constant $\alpha(k)>0$, then taking $v=u(k)-u_M$ and
using $A(k)(u(k)-u_M)=F-A(k)u_M$ gives the error bound:
\begin{equation}
  \|u(k)-u_M\|_{K_0}\;\le\;\frac{\|F-A(k)\,u_M\|_{K_0^{-1}}}{\alpha(k)} .
  \label{eq:cert}
\end{equation}
Eq.~(\ref{eq:cert}) is the standard residual bound of reduced-basis methods
\citep{rozza2008reduced}, where the residual is computed for the parametrized problem. In INO, it can be evaluated for either field: with the truncated field
of Eq.~(\ref{eq:kl}) it bounds the rank error, and with the true input field it bounds the total
error. The bound requires only the residual $F-A(k)u_M$ and the constant $\alpha(k)$, both of which
are computed from the values of $k$ at the quadrature points. Therefore, it requires no reference
solution, holds outside the parametric domain whenever $\alpha(k)>0$, and applies to any $u_M$ that
satisfies the Dirichlet boundary conditions, including the prediction of a trained neural operator
(Section~\ref{sec:accuracy}). In this paper, $K_0$ is the stiffness matrix of the Laplacian, $\int_\Omega\nabla v\cdot\nabla u$,
which does not depend on the input field, so that $\|\cdot\|_{K_0}$ is the energy norm and
$\alpha(k)$ has a closed form. For diffusion, $\alpha(k)=\min_q k(\bm x_q)$ is the minimum
of the coefficient over the quadrature points $\bm x_q$. For the Helmholtz problem
$-\Delta u-\kappa^2u=f$ with the squared wavenumber $\kappa^2$ as input,
$\alpha(\kappa^2)=1-\kappa^2_{\max}/\mu_1$, where $\kappa^2_{\max}$ is the largest value of
$\kappa^2$ at the quadrature points and $\mu_1$ is the first Dirichlet eigenvalue of the Laplacian,
so that $\alpha>0$ as long as $\kappa^2_{\max}<\mu_1$. If $\alpha\le0$, the bound is not defined. The bound
measures the error with respect to the discrete solution $u(k)$, so it does not include the
discretization error; accordingly, all reference solutions in this paper except that of the
moving source use the same discretization as INO.

\section{Results}

In this section, we compare INO with physics-informed neural operators. All comparisons with the baselines follow the same protocol. Each
test input is a sample of the random field with all modes retained, each method receives this true
input field, and the reference solution is computed for the true input field, so that the errors
in Tables~\ref{tab:neural} and~\ref{tab:alldraw} are total errors. The baselines are trained for the number of iterations given in their original papers
\citep{li2021physics,wang2021learning}, and all errors are those of the fully trained models. Unless stated otherwise,
errors are relative $L^2$ errors evaluated by Gaussian quadrature at the same points for every
method. We also report the error of the mean-input solution, i.e., the solution for the mean input
field, as the error of a model that ignores its input (Table~\ref{tab:alldraw}). The PDEs, input distributions, reference solutions and detailed settings
are given in Appendix~\ref{app:setup}. All INO timings are measured on one CPU core.

\subsection{Comparison with physics-informed operators}
\label{sec:neural}

\begin{table}[t]
\caption{Relative $L^2$ errors (for the moving source, relative $\ell^2$ errors on $500\times51$
points in $(x,t)$) of INO and the physics-informed baselines on the test samples within the
parametric domain of INO, selected as described in Section~\ref{sec:neural} (all samples:
Table~\ref{tab:alldraw}). Each cell gives the geometric mean over three random initializations and,
in parentheses, the ratio of the largest to the smallest error; a cell without parentheses is a
single initialization, and \textbf{bold} marks
the most accurate model. Every model is trained on the mesh of the problem. Training time
is given in seconds as INO / PI-FNO / PI-DeepONet, with INO on one CPU core and the baselines on one
GPU. The settings of the baselines are given in Appendix~\ref{app:baselines}.}
\label{tab:neural}
\begin{center}
\footnotesize\setlength{\tabcolsep}{3pt}
\begin{tabular}{lllll}
\toprule
problem (mesh) & \INO & PI-FNO & PI-DeepONet & training time (s) \\
\midrule
1D heat, moving source      & \NUM{$\bm{2.92}$e-3 (1.13)} & \NUM{3.98e-1 (1.25)} & \NUM{3.89e-1 (1.91)} & \NUM{115} / \NUM{5{,}784} / \NUM{361} \\
1D diffusion--reaction      & \NUM{$\bm{4.84}$e-5 (1.41)} & \NUM{2.56e-3 (1.35)} & \NUM{5.21e-3 (1.04)} & \NUM{8} / \NUM{53} / \NUM{202} \\
2D diffusion $65^2$         & \NUM{$\bm{4.34}$e-3 (1.01)} & \NUM{5.53e-3 (1.45)} & \NUM{5.42e-2 (1.07)} & \NUM{42} / \NUM{1{,}718} / \NUM{2{,}016} \\
2D diffusion $129^2$        & \NUM{$\bm{4.23}$e-3 (1.09)} & \NUM{4.81e-2 (1.18)} & \NUM{5.85e-2 (1.13)} & \NUM{79} / \NUM{6{,}189} / \NUM{2{,}047} \\
2D diffusion $513^2$        & \NUM{$\bm{4.58}$e-3 (1.08)} & \NUM{1.63e-1 (1.02)} & \NUM{5.25e-2 (1.04)} & \NUM{530} / \NUM{77{,}231} / \NUM{1{,}844} \\
2D Helmholtz $65^2$         & \NUM{$\bm{4.14}$e-4 (1.07)} & \NUM{6.17e-3 (1.77)} & \NUM{1.04e-1 (1.45)} & \NUM{22} / \NUM{1{,}768} / \NUM{1{,}973} \\
2D Helmholtz $129^2$        & \NUM{$\bm{4.17}$e-4 (1.08)} & \NUM{6.70e-2 (1.72)} & \NUM{1.17e-1 (1.24)} & \NUM{35} / \NUM{6{,}229} / \NUM{1{,}960} \\
2D Helmholtz $513^2$        & \NUM{$\bm{3.74}$e-4 (1.21)} & \NUM{7.99e-2 (1.00)} & \NUM{9.93e-2 (1.09)} & \NUM{373} / \NUM{77{,}627} / \NUM{1{,}700} \\
3D diffusion $65^3$         & \NUM{7.67e-3 (1.02)} & \NUM{$\bm{2.08}$e-3 (1.53)} & \NUM{2.08e-1 (1.09)} & \NUM{452} / \NUM{15{,}049} / \NUM{2{,}750} \\
3D diffusion $129^3$        & \NUM{8.73e-3 (1.12)} & \NUM{$\bm{6.57}$e-3  (1.15)} & \NUM{2.00e-1 (1.06)} & \NUM{20{,}499} / \NUM{102{,}879} / \NUM{3{,}269} \\
3D Helmholtz $65^3$         & \NUM{$\bm{5.88}$e-4 (1.08)} & \NUM{1.20e-3 (1.82)} & \NUM{4.07e-1 (1.36)} & \NUM{3{,}335} / \NUM{15{,}095} / \NUM{2{,}405} \\
3D Helmholtz $129^3$        & \NUM{$\bm{6.62}$e-4 (1.08)} & \NUM{9.35e-4  (1.73)} & \NUM{4.16e-1 (1.26)} & \NUM{3{,}053} / \NUM{103{,}236} / \NUM{3{,}051} \\
2D nonlinear diffusion $65^2$& \NUM{$\bm{2.79}$e-3 (1.06)} & \NUM{9.11e-3 (2.25)} & \NUM{2.21e-1 (1.05)} & \NUM{278} / \NUM{813} / \NUM{2{,}306} \\
2D nonlinear diffusion $129^2$& \NUM{$\bm{2.58}$e-3 (1.10)} & \NUM{5.28e-2 (4.46)} & \NUM{2.21e-1 (1.17)} & \NUM{348} / \NUM{2{,}194} / \NUM{2{,}345} \\
\bottomrule
\end{tabular}
\end{center}
\end{table}

In this example, we compare INO with PI-FNO and PI-DeepONet on the moving-source,
diffusion--reaction, diffusion, Helmholtz and nonlinear diffusion problems of
Appendix~\ref{app:problems}. Each baseline is trained with the settings of its original paper
(Appendix~\ref{app:baselines}). The two-dimensional operators of INO use $S=25$ KL coordinates,
$M=96$ modes and a parametric domain of $\pm2$ standard deviations, and the three-dimensional
operators use up to $S=164$ coordinates (Table~\ref{tab:setup}). The moving-source,
diffusion--reaction and nonlinear diffusion operators use Hermite parametric factors and have no
bounded parametric domain.

INO makes a prediction only if the six leading coordinates of the input, i.e., the coordinates of the
six largest KL eigenvalues, lie within the parametric domain; we call this test the domain check.
The remaining coordinates are clipped to the parametric domain. Only the leading coordinates are
checked, since the trailing KL modes have small amplitudes, so clipping them hardly changes the input
field (Appendix~\ref{app:accepted}). For a Gaussian input field, the
probability of passing the check is $0.9545^6\approx76\%$, and \NUM{71\%} of our two-dimensional
samples pass it. In three dimensions, the check uses the ten leading coordinates and keeps \NUM{70\%}
of the samples. Since the moving-source and diffusion--reaction operators have no bounded
parametric domain, they are evaluated on every sample. The nonlinear diffusion operator has none
either, but it is evaluated on the samples that pass the same check (\NUM{77\%}), so that all two-dimensional problems are compared on the same kind of samples;
on every sample, its error is about \NUM{10\%} larger. All models in Table~\ref{tab:neural} are evaluated on the same samples,
and the results on every sample are given in Appendix~\ref{app:accepted}. Fig.~\ref{fig:fields}
shows three predictions of INO, each for the test sample with the median error among the samples of
Table~\ref{tab:neural}.

\begin{figure}[t]
\begin{center}
\includegraphics[width=0.97\linewidth]{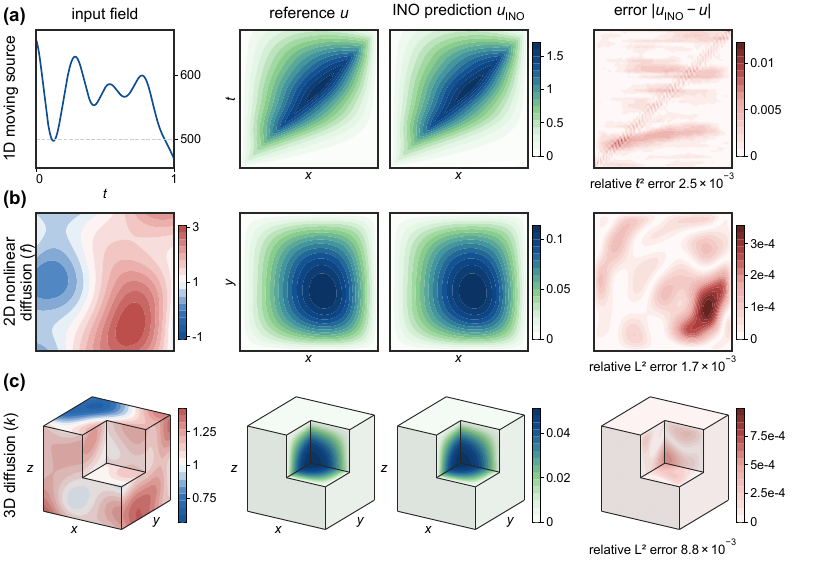}
\end{center}
\caption{Predictions of INO. (a) The moving heat source on the $2001\times51$ space--time mesh for a
Gaussian sample of the power history $P(t)$, with the operator of Table~\ref{tab:neural}; the
reference is a Crank--Nicolson solution, and the relative $\ell^2$ error is
\NUM{$2.5\times10^{-3}$}. (b) Two-dimensional nonlinear diffusion at $129^2$ with the operator of
Table~\ref{tab:neural}, for a Gaussian sample of the source $f$; the reference is a Newton solve for
the true source, and the relative error is \NUM{$1.7\times10^{-3}$}. (c)
Three-dimensional diffusion at $129^3$ with the operator of Table~\ref{tab:neural}, for a Gaussian
sample of the coefficient $k$; the reference is solved for the true input field, and the relative
error is \NUM{$8.8\times10^{-3}$}. }
\label{fig:fields}
\end{figure}

INO is the most accurate model in 12 of the 14 cases of Table~\ref{tab:neural}, and each factor
below is relative to the more accurate baseline. On
two-dimensional Helmholtz, it is \NUM{15}--\NUM{214}$\times$ more accurate than PI-FNO, and on
two-dimensional diffusion, it is \NUM{1.3}$\times$ more accurate at $65^2$ and \NUM{11}$\times$ at
$129^2$ and $513^2$. It is also more accurate on three-dimensional Helmholtz, by \NUM{2.0}$\times$ at
$65^3$ and \NUM{1.4}$\times$ at $129^3$, and on the nonlinear diffusion problem, by
\NUM{3.3}$\times$ at $65^2$ and \NUM{20}$\times$ at $129^2$. On the moving-source problem, both
baselines have larger errors than the mean-input solution (\NUM{0.13}). PI-FNO is more accurate on three-dimensional diffusion, by \NUM{3.7}$\times$ at $65^3$ and \NUM{1.3}$\times$ at
$129^3$.

The 1D diffusion--reaction problem is the space--time benchmark of \citet{wang2021learning},
$\partial_t u=0.01\,\partial_{xx}u+0.01\,u^2+f(x)$ for $(x,t)\in(0,1)\times(0,1]$ with zero
initial and boundary conditions, where the source $f$ is the input field. Since the reaction term is
a polynomial in $u$, INO linearizes it about the previous iterate, and each linearized problem keeps
the separated form with a number of terms that is independent of $S$. On this problem, INO is
\NUM{53}$\times$ more accurate than PI-FNO, whereas the mean-input solution, which is zero since the source has zero
mean, has an error of \NUM{1.00}.

The error of INO is nearly independent of the mesh, whereas the error of PI-FNO trained on the
$129^2$ and $513^2$ grids is \NUM{6}--\NUM{30}$\times$ larger than at $65^2$, since its training loss
stops decreasing within a few thousand steps on the finer grids. If the network trained at $65^2$ is applied
on the finer grids instead, PI-FNO is more accurate than INO on two-dimensional diffusion, but not on
Helmholtz or nonlinear diffusion (Appendix~\ref{app:finemesh}). PI-DeepONet
trains quickly, but on the moving-source, Helmholtz and three-dimensional problems its error is
larger than that of the mean-input solution (Appendix~\ref{app:baselines}). The error of PI-FNO changes by \NUM{1.0}--\NUM{4.5}$\times$ across random initializations, whereas that of INO
changes by at most \NUM{1.4}$\times$ (Appendix~\ref{app:variability}).

\begin{figure}[t]
\begin{center}
\includegraphics[width=0.95\linewidth]{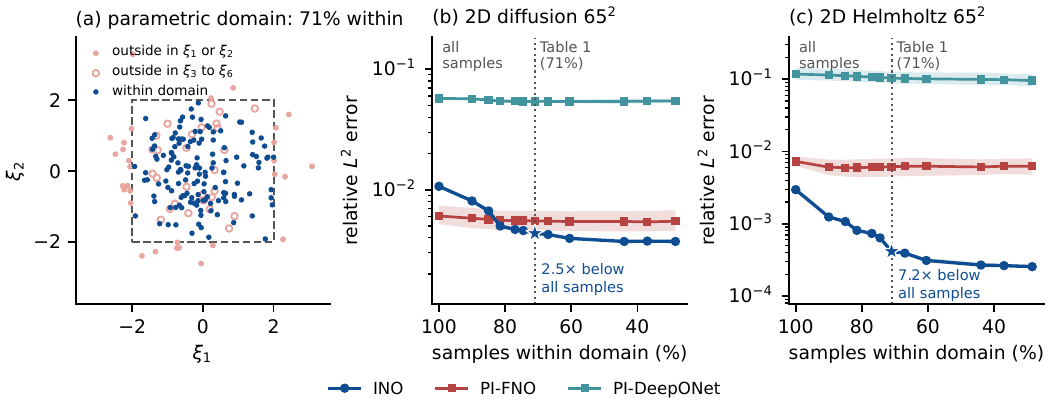}
\end{center}
\caption{Accuracy versus the fraction of samples within the parametric domain at $65^2$. (a) The
\NUM{200} diffusion samples in the plane of their two leading coordinates $\xi_1$ and $\xi_2$ and the parametric domain
(dashed); since the check uses the six leading coordinates, the hollow samples lie inside this
square but outside the parametric domain in $\xi_3$ to $\xi_6$.
(b, c) Error versus the fraction of samples within the parametric domain as the check is tightened,
with every model evaluated on the same samples (geometric mean and range over three
initializations). The star marks the setting of Table~\ref{tab:neural}, and the left end of each
curve includes every sample.}
\label{fig:domain}
\end{figure}

As shown in Fig.~\ref{fig:domain}, tightening the domain check from every sample to the setting of
Table~\ref{tab:neural} reduces the error of INO at $65^2$ by \NUM{2.5}$\times$ on diffusion and
\NUM{7.2}$\times$ on Helmholtz, whereas the errors of PI-FNO and of the mean-input solution change by
less than \NUM{30\%}, since INO is trained only on the parametric domain. Most of the error of INO thus comes from the inputs outside the parametric domain, which the domain check
identifies from the input before any prediction. We use the check as a coverage policy, in which INO
predicts the inputs within its parametric domain and the other inputs are passed to a conventional
solver or to an INO trained on a wider parametric domain, as long as the coefficient remains
positive over it (Section~\ref{sec:query}). A rejected input is not necessarily predicted poorly,
since the check only tests whether the input lies within the domain on which INO is trained. When every sample is included, INO is still
\NUM{2.5}--\NUM{32}$\times$ more accurate on two-dimensional Helmholtz and remains the most accurate
model on two-dimensional diffusion at $129^2$ and $513^2$ and on the nonlinear diffusion problem,
whereas PI-FNO is the most accurate model on two-dimensional diffusion at $65^2$, by
\NUM{1.8}$\times$, and on every three-dimensional problem (Appendix~\ref{app:accepted}).

\subsection{Rank error and error bound}
\label{sec:accuracy}

\begin{figure}[t]
\begin{center}
\includegraphics[width=0.93\linewidth]{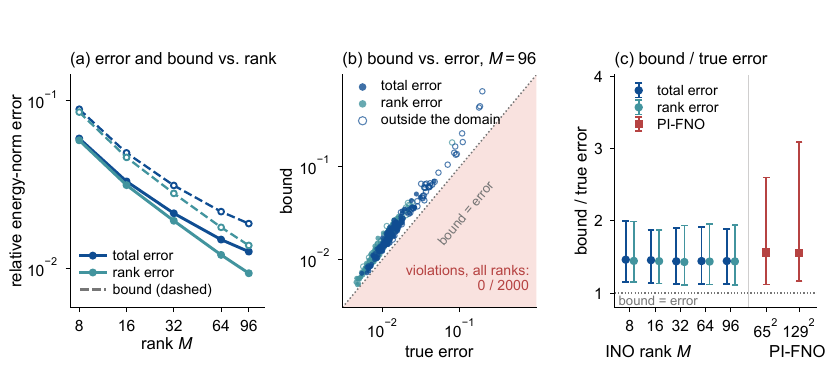}
\end{center}
\caption{Error bound of Eq.~(\ref{eq:cert}) for the two-dimensional diffusion operator of
Table~\ref{tab:neural} at $65^2$ ($S=25$, parametric domain of $\pm2$ standard deviations), on the
\NUM{200} test samples of the full random field, \NUM{71\%} of which lie within the parametric
domain. Errors are relative energy-norm errors; the relative $L^2$ error of the
same predictions at $M=96$ is \NUM{$4.3\times10^{-3}$} (Table~\ref{tab:neural}); blue: total error
with respect to the true input field, teal: rank error with respect to the truncated field. (a) Mean
true error (solid) and mean bound (dashed) versus rank, samples within the domain.
(b) Bound versus true error for every sample at $M=96$; the shaded region would contain violations. (c) Ratio of the
bound to the true error (mean, minimum and maximum over the samples within the domain) for INO at
every rank and for the predictions of PI-FNO trained at $65^2$ and applied at $65^2$ and $129^2$.}
\label{fig:bound}
\end{figure}

In this example, we study how the error of INO decreases with the rank $M$ and how closely the bound
of Eq.~(\ref{eq:cert}) follows it. We use the two-dimensional diffusion operator of
Table~\ref{tab:neural} at $65^2$ ($S=25$) and the test samples within its parametric domain. In the relative energy norm, the total error decreases from \NUM{$6.0\times10^{-2}$} at $M=8$ to
\NUM{$1.3\times10^{-2}$} at $M=96$, and the rank error from \NUM{$5.8\times10^{-2}$} to
\NUM{$9.4\times10^{-3}$} (Fig.~\ref{fig:bound}). The gap between the two
errors is caused by the truncation of the input field, so it does not decrease with $M$.

The bound is never violated in \NUM{2{,}000} evaluations, and it is \NUM{1.43}--\NUM{1.46} times the
true error at every rank, for both the total error and the rank error. The bound thus follows the true error closely, and the rank beyond which adding modes barely reduces the total error can be
read from the bound alone, without any reference solution. The rank error also hardly depends on the mesh, changing by less than a factor of two from $65^2$
to $513^2$ for $S=9$ and $M=96$ (Appendix~\ref{app:rankmesh}).

Since Eq.~(\ref{eq:cert}) only requires a prediction that satisfies the Dirichlet boundary
conditions, it can also be applied to the baselines. PI-FNO satisfies them exactly, since its output
is multiplied by a mollifier that vanishes on the boundary (Appendix~\ref{app:baselines}). For PI-FNO
on two-dimensional diffusion, trained
at $65^2$ and applied at $65^2$ and $129^2$, the bound is never violated in \NUM{240} predictions, and
its mean ratio to the true energy-norm error is \NUM{1.5}--\NUM{1.6} (Fig.~\ref{fig:bound}(c)). On
the other hand, the predictions of PI-DeepONet, whose boundary conditions are imposed only by a
penalty term, violate the boundary condition by about \NUM{12\%} of the solution maximum, so the bound does not apply to them and is violated in \NUM{117} of \NUM{120}
cases. After the boundary values are imposed, the bound holds again.

\subsection{Efficiency and error analysis}
\label{sec:efficiency}

As shown in Table~\ref{tab:neural}, training INO takes \NUM{22}--\NUM{530}\,s
on the two-dimensional problems, compared with \NUM{813}--\NUM{77{,}627}\,s for PI-FNO and
\NUM{1{,}700}--\NUM{2{,}345}\,s for PI-DeepONet on one GPU. On the three-dimensional problems at
$129^3$, it takes \NUM{3{,}053}--\NUM{20{,}499}\,s, compared with \NUM{102{,}879}--\NUM{103{,}236}\,s
for PI-FNO and \NUM{3{,}051}--\NUM{3{,}269}\,s for PI-DeepONet. The training time of INO grows much more slowly than the number of degrees of freedom. From $65^2$
to $513^2$, the degrees of freedom increase by a factor of \NUM{62} and the training time by a factor of \NUM{13} to \NUM{17},
since mesh refinement only affects $d$ of the $d+S$ directions (Appendix~\ref{sec:dsweep}).

INO stores $M(d\,n+S\,n_\xi)$ numbers, from \NUM{91{,}520} for the $S=8$, $M=128$ operator of
Table~\ref{tab:accuracy} to about \NUM{2.2} million for the $S=164$ operators of
Table~\ref{tab:neural} at $129^3$, which is about the number of nodal values of a single solution on
that mesh ($129^3\approx2.1$ million). The inference cost is compared with a conventional numerical
solve in Table~\ref{tab:query-cost} (Appendix~\ref{sec:cost}). The speed-up
increases from \NUM{484}$\times$ at $65^2$ to \NUM{2{,}892}$\times$ at $513^2$, since an inference
only contracts one-dimensional factors, whereas the cost of a numerical solve grows faster than the
number of degrees of freedom. Against a matrix-free solve of the same C-HiDeNN discretization, the
speed-up at $65^2$ is \NUM{375}$\times$ for $M=64$. For an input given by its KL coordinates, an
inference together with its total-error bound costs about half of a numerical solve at $65^2$ and a
quarter at $129^2$, and for a full random-field input, the bound costs \NUM{0.2}--\NUM{0.5}\,s
(Appendix~\ref{sec:cost}).

We further examine the source of the error of INO. The prediction is a sum of spatial modes
$\prod_i u^m_{x_i}$ weighted by modal coefficients $\prod_j u^m_j(\xi^\star_j)$, which depend on
the input. At $65^2$ with $S=9$ and $M=96$, the best approximation of the reference solution in the space
spanned by the spatial modes has a relative error of \NUM{$1.4\times10^{-5}$}, whereas the error of
INO is \NUM{$3.6\times10^{-4}$}. Therefore, the error comes from the modal coefficients rather than
from the spatial modes. Since the matrix $A(k)$ is affine in $\bm\xi$ for the coefficient inputs, the
modal coefficients can instead be computed for a given input by solving a reduced system of size
$r\le M$. This reduces the error by one to two orders of magnitude on the linear problems, at an
inference cost of about one millisecond or less (Appendix~\ref{sec:readoff}).

\section{Limitations}
\label{sec:limitations}

The error of INO is limited by the truncation of the input field rather than by the rank
(Section~\ref{sec:accuracy}). The
parametric domain of $\pm2$ standard deviations excludes \NUM{29\%} of the samples, which require a
conventional numerical solve or an INO parameterized on a wider parametric domain. A coefficient input must remain
positive over the whole parametric domain, which limits its contrast, whereas source and reaction inputs have
no such limit. The input fields must also be smooth,
and the bound does not cover the discretization error.

\section{Conclusion}
\label{sec:conclusion}

In this paper, we introduced INO, a data-free interpolating neural network for the solution
operators of parametric PDEs, which is trained by one-dimensional linear solves and whose total error
is bounded by the residual for the true input field on coercive problems. INO is the most accurate
model on most of the problems within its parametric domain, whereas the physics-informed FNO is more
accurate on three-dimensional diffusion. In future work, we will use the computable bound to adapt the rank and the parametric domain during
training, extend INO to transport-dominated and multiphysics problems, and use its certified
predictions for design optimization, inverse problems and uncertainty quantification.

\subsection*{AI use statement}
In this work, we used generative AI tools for implementing the baseline Crank-Nicolson finite difference solvers. We have not used generative AI tools for generating synthetic data sets, helping develop theoretical models or conceptual frameworks, formulating mathematical claims, providing critical ingredients for proving mathematical claims, assisting in the writing of proofs, proposing or refining hypotheses, designing or providing feedback on research methodology or experiments, assisting with translation, cleaning or reformatting datasets, supporting qualitative and thematic data analysis, or interpreting results. Additionally, we used generative AI tools for tasks such as editing a research paper to improve readability, identifying relevant literature, and formatting references. We have reviewed all AI-assisted work. We take responsibility for the final content of this work, including text, claims
or artifacts produced with the aid of generative AI.

\subsection*{Ethics statement}
This work develops AI-enhanced numerical methods for parametric partial differential equations to approximate PDE operators. We are not aware of ethical concerns specific to
this work beyond those that apply to scientific computing in general.

\subsection*{Reproducibility statement}
Section~\ref{sec:method} and Appendices~\ref{app:chidenn} and~\ref{app:algebra} describe the
network, the training step and the inference of INO, including the one-dimensional systems solved
during training. Appendix~\ref{app:problems} defines every problem, input distribution and
reference solver; Appendix~\ref{app:setup} lists the discretization, rank, parametric domain and
solver settings of every experiment (Table~\ref{tab:setup}); and Appendix~\ref{app:baselines} gives
the architectures, training budgets and evaluation of the baselines, which follow their published
implementations. All results use fixed random seeds and the same test samples for every model, and
the variability over initializations is reported in Appendix~\ref{app:variability}. 

\bibliographystyle{plainnat}
\bibliography{references}

\appendix
\section{C-HiDeNN interpolation and structure of INO}
\label{app:chidenn}
In this section, we describe the C-HiDeNN sub-network used for the factors of INO and how INO
combines the sub-networks.

\subsection{C-HiDeNN sub-network}
\begin{figure}[t]
\begin{center}
\includegraphics[width=\linewidth]{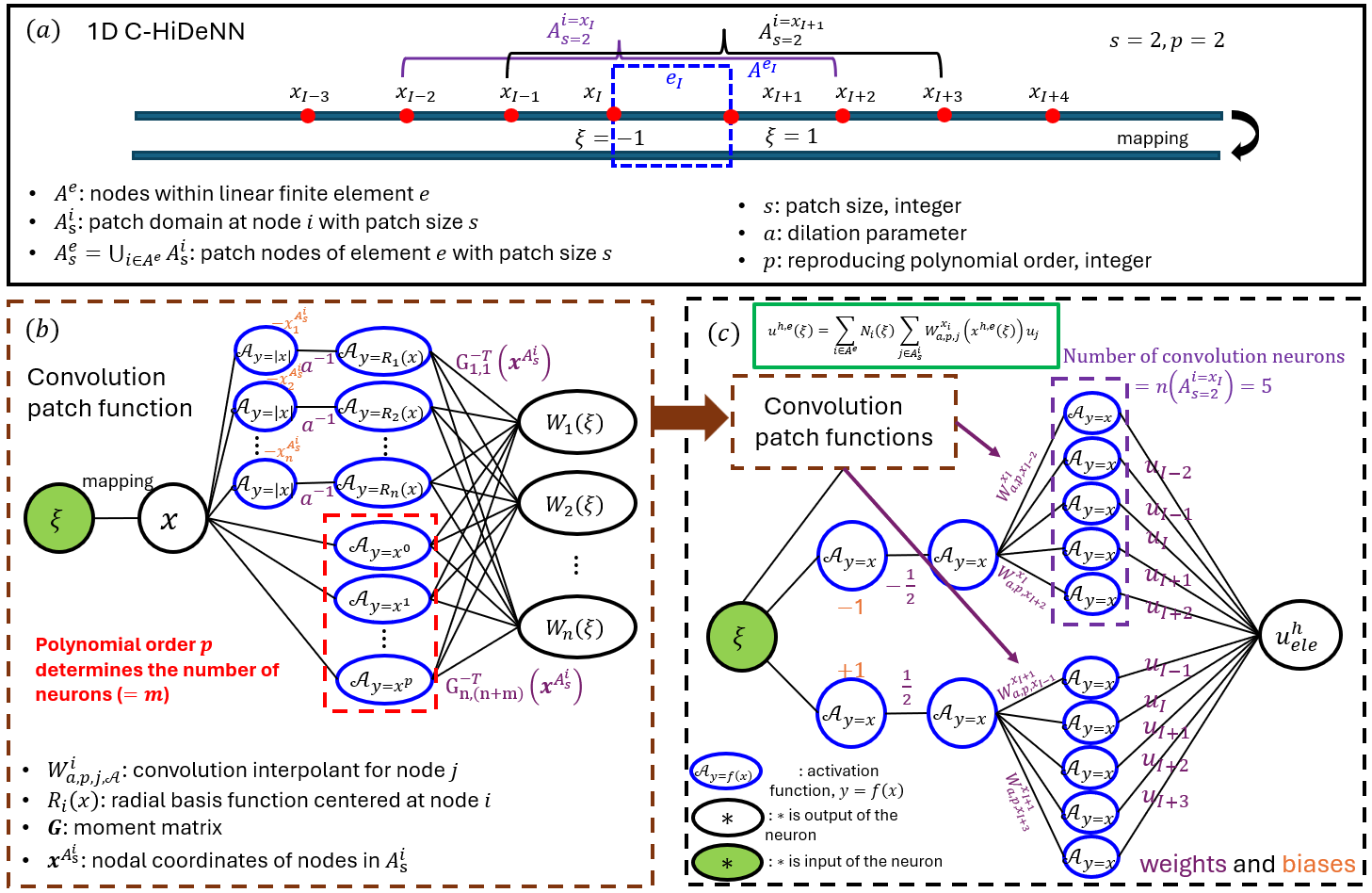}
\end{center}
\caption{C-HiDeNN interpolation of a one-dimensional factor interpreted as a neural network.
(a) Element, patch and hyperparameters $s$, $a$, $p$. (b) Convolution patch function formed from
radial basis and polynomial neurons. (c) The factor as a partially connected network whose output is
contracted with the nodal values. Adapted from \citet{park2024engineering}.}
\label{fig:chidenn}
\end{figure}

Convolution
hierarchical deep-learning neural network (C-HiDeNN) interpolation combines finite element
interpolation, meshfree interpolation and machine learning \citep{lu2023convolution,park2023convolution}.
A one-dimensional C-HiDeNN interpolation can be written as:
\begin{equation}
  u(x)=\sum_{I\in A^e}N_I(x)\sum_{J\in A_s^I}W^{(J)}_I(x;s,a,p)\,u_J
      =\sum_{J\in A_s^e}\widetilde N_J(x;s,a,p)\,u_J=\widetilde{\bm N}(x)\,\bm u ,
  \label{eq:chidenn}
\end{equation}
where $N_I(x)$ is the linear finite element basis function of node $I$ of element $e$; $A^e$ is the
set of nodes of the element; $W^{(J)}_I(x;s,a,p)$ is the convolution patch function of node $J$ on
the patch $A_s^I$ of the $2s{+}1$ nodes around node $I$; $A_s^e=\bigcup_{I\in A^e}A_s^I$ is the set of
patch nodes of the element; and $u_J$ is the nodal value at node $J$. As shown in
Fig.~\ref{fig:chidenn}, Eq.~(\ref{eq:chidenn}) can be interpreted as a partially connected neural
network whose hidden layers produce the functions $N_I$ and $W^{(J)}_I$ and whose only trainable
parameters are the nodal values $u_J$. The patch functions are controlled by the patch size $s$, the
dilation parameter $a$ and the reproducing order $p$, so the function space can be adapted without
changing the number of nodes.

On each element, the C-HiDeNN basis functions can be written as:
\begin{equation}
\begin{gathered}
  \widetilde N_J(x)=\sum_{I\in A^e}N_I(x)\,W^{(J)}_{I}(x),\\
  \bm W^{I}(x)^{\top}\ =\big[\bm R^{I}(x)^{\top}\ \ \bm p(x)^{\top}\big]\,\mathbf G_I^{-1},\qquad
  \mathbf G_I=\begin{bmatrix}\mathbf R_0 & \mathbf P\\ \mathbf P^{\top} & \mathbf 0\end{bmatrix},
\end{gathered}
\end{equation}
where $\bm W^{I}(x)$ collects the patch functions $W^{(J)}_I(x)$ of the nodes $J\in A_s^I$;
$\mathbf R_0=[R(x_J-x_{J'})]$ and $\mathbf P=[\bm p(x_J)^\top]$ run over the nodes $J,J'$ of that
patch; $\bm p(x)=[1,x,\dots,x^{p}]$; $\bm R^{I}(x)=[R(x-x_J)]$; and the entries of the product
associated with the polynomial constraints are not used. The radial basis function is the
cubic spline $R(z)=\tfrac23-4z^2+4z^3$ for $0\le z\le\tfrac12$,
$R(z)=\tfrac43-4z+4z^2-\tfrac43z^3$ for $\tfrac12<z\le1$, and zero otherwise, with $z=|x-x_J|/a$,
where $a$ is a multiple of the element size $h$ (Appendix~\ref{app:setup}).

The C-HiDeNN basis functions are interpolatory and subject to the Kronecker delta property, i.e., $\widetilde N_I(x_J)=\delta_{IJ}$. As a result,
the nodal values are the values of the function at the nodes, and a homogeneous Dirichlet condition
is imposed exactly by removing the boundary unknowns. The basis functions also reproduce polynomials up to degree $p$. For a two-dimensional diffusion problem with a smooth random
coefficient, the $L^2$ error of C-HiDeNN with $s=3$ and $p=2$ converges with order \NUM{3.44},
compared with \NUM{2.00} for linear elements, and it is \NUM{72}$\times$ smaller at $65^2$.
Therefore, the discretization error stays far below the rank error in all experiments.

\subsection{Structure of INO}
\label{app:structure}
\begin{figure}[t]
\begin{center}
\includegraphics[width=0.9\linewidth]{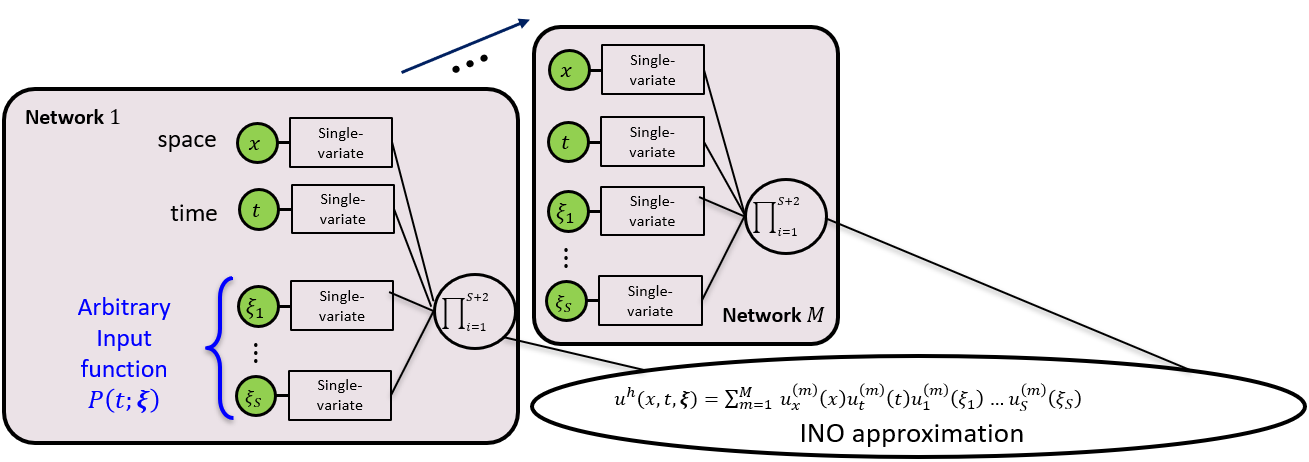}
\end{center}
\caption{Structure of INO for the moving heat source. Each of the $M$ networks has one
single-variate sub-network per input, namely $x$, $t$ and the KL coordinates of the power history
$P$; the outputs of the sub-networks are multiplied, and the $M$ networks are summed, as in
Eq.~(\ref{eq:ansatz-move}).}
\label{fig:inn}
\end{figure}

We use the moving heat source of Appendix~\ref{app:problems} to show how INO combines the
sub-networks. The input of this problem is the power history $P(t)$,
which is represented by its first $S=18$ KL coordinates $\xi_1,\dots,\xi_S$, and the solution operator
maps $P$ to the space--time field $u(x,t)$. INO approximates this operator as:
\begin{equation}
  u(x,t;\bm\xi)\;\approx\;\sum_{m=1}^{M}u^{m}_{x}(x)\,u^{m}_{t}(t)\prod_{j=1}^{S}u^{m}_{j}(\xi_j),
  \label{eq:ansatz-move}
\end{equation}
where $u^m_x$, $u^m_t$ and $u^m_j$ are the single-variate sub-networks of the $m$-th mode in $x$, $t$
and $\xi_j$. As shown in Fig.~\ref{fig:inn}, each mode is a network with one sub-network per input;
the outputs of the sub-networks are multiplied, and the $M$ networks are summed. The sub-networks in
$x$ and $t$ are C-HiDeNN interpolants of Eq.~(\ref{eq:chidenn}), and the parametric sub-networks of
this problem are the Hermite expansions of Appendix~\ref{app:problems}. For the other problems, the
inputs $x$ and $t$ are replaced by the $d$ spatial coordinates, and every sub-network is a C-HiDeNN
interpolant, which gives Eq.~(\ref{eq:ansatz}).

\section{Separated form, training step and inference}
\label{app:algebra}
\label{sec:rule}
In this section, we derive the separated form of the weak form for two-dimensional diffusion, the
one-dimensional training step of Section~\ref{sec:boosting} and the inference of
Section~\ref{sec:query}.

\subsection{Separated form}
Sans-serif letters denote one-dimensional matrices, and $\mathsf K$ and $\mathsf M$ are the stiffness
and mass matrices $\int\widetilde{\bm N}'^{\top}\widetilde{\bm N}'$ and
$\int\widetilde{\bm N}^{\top}\widetilde{\bm N}$ of a one-dimensional C-HiDeNN mesh. Since $k$ is
affine in $\bm\xi$, substituting Eq.~(\ref{eq:kl}) into the bilinear form $\mathcal A_k$ of
$-\nabla\!\cdot(k\nabla u)=f$ splits it, in one spatial dimension, into exactly $S{+}1$ separated
terms, which can be written as:
\begin{equation}
  \int k(\bm x;\bm\xi)\,\nabla\delta u\cdot\nabla u \,\mathrm d\bm x\,\mathrm d\bm\xi
  \;=\;\underbrace{\bar k\,\mathsf K\otimes\bigotimes_{j'} \mathsf M_{j'}}_{\text{mean}}
  \;+\;\sum_{j=1}^{S}\underbrace{\mathsf K^{(j)}\otimes \mathsf M^{\xi}_{j}\otimes\bigotimes_{j'\neq j}\mathsf M_{j'}}_{\text{one weighted direction}} ,
  \label{eq:split}
\end{equation}
where $\mathsf K^{(j)}=\int\phi_j\,\widetilde{\bm N}'^{\top}\widetilde{\bm N}'$ is the stiffness matrix
weighted by the KL mode $\phi_j$; $\mathsf M_{j}$ is the mass matrix of parameter direction $j$; and
$\mathsf M^\xi_{j}=\int\widetilde{\bm N}^\top\xi\,\widetilde{\bm N}\,\mathrm d\xi$ is its
$\xi$-weighted mass matrix.

In two dimensions, the covariance is a product over the spatial directions, so every KL mode is a
product of one-dimensional functions, $\phi_j(x,y)=\phi^x_j(x)\,\phi^y_j(y)$. Each term of
Eq.~(\ref{eq:split}) then splits into one term per gradient component, whose matrix in $x$ is the
stiffness or mass matrix weighted by $\phi^x_j$ and whose matrix in $y$ is the other of the two,
weighted by $\phi^y_j$ (both unweighted for the mean term). This gives $T=2(S{+}1)$ separated terms in two dimensions and $T=d(S{+}1)$ in
$d$ dimensions, so the matrix of the bilinear form can be written as:
\begin{equation}
  A=\sum_{t=1}^{T}\gamma_t\,\mathsf A^t_x\otimes\mathsf A^t_y\otimes\bigotimes_j\mathsf P^t_j ,
  \label{eq:terms}
\end{equation}
where $\gamma_t$ is the scalar coefficient of term $t$ ($\bar k$ for the mean terms and $1$
otherwise); $\mathsf A^t_x$ and $\mathsf A^t_y$ are its one-dimensional spatial matrices; and
$\mathsf P^t_j\in\{\mathsf M_j,\mathsf M^\xi_j\}$ is its matrix in parameter direction $j$.

\subsection{Training step}
For every problem, INO is trained on the Galerkin weak form of Section~\ref{sec:boosting}. Modes are
added one at a time, and the $M$-th mode $w_M$ is the rank-one sub-network that satisfies the weak
form for the residual left by the previous modes, which can be written as:
\begin{equation}
  \int_{\Omega_{\bm\xi}}\mathcal A_k\big(\delta w,\,u_{M-1}+w_M\big)\,\mathrm d\bm\xi
  \;=\;\int_{\Omega_{\bm\xi}}F(\delta w)\,\mathrm d\bm\xi ,
  \qquad
  u_{M} \;=\; \sum_{m=1}^{M} c_m\,w_m ,
  \label{eq:boost}
\end{equation}
where $\delta w$ is any variation of one factor of $w_M$; $w_m$ is the $m$-th rank-one mode; $c_m$ is
its amplitude; and $u_M$ is the prediction with $M$ modes. For a symmetric operator,
Eq.~(\ref{eq:boost}) is the stationarity condition of the energy functional
$\mathcal J(u)=\int_{\Omega_{\bm\xi}}\big[\tfrac12\mathcal A_k(u,u)-F(u)\big]\,\mathrm d\bm\xi$, so
every new mode decreases $\mathcal J$, as in the gradient boosting of \citet{friedman2001greedy}. If
$\delta w$ varies only one factor of the new mode, every other direction contributes a scalar to each
term, so the $T$ terms reduce to a weighted sum of one-dimensional matrices. For the $x$ direction of
the new mode $v_x\otimes v_y\otimes\bigotimes_j v_j$, the resulting linear system can be written as:
\begin{equation}
  \Big[\textstyle\sum_{t} \gamma_t\, \theta_t\; \mathsf A^{t}_{x}\Big]\, v_x \;=\; \bm q ,
  \label{eq:onedsolve}
\end{equation}
where $\theta_t=(v_y^\top\mathsf A^t_yv_y)\prod_j(v_j^\top\mathsf P^t_jv_j)$ collects the inner products
of the fixed factors for term $t$; and $\bm q$ contains the load and the previous modes through the
same products. Therefore, each step is one dense solve of size $n$, the cost grows linearly with $S$,
and no matrix of the size of the spatial grid is assembled. After each enrichment, the amplitudes
$c_m$ of all modes are updated by a Galerkin projection.

\subsection{Inference}
For a new input field $k^\star$, the field is projected onto the KL modes, and the network is
evaluated at the resulting coordinates, which can be written as:
\begin{equation}
  \bm\xi^\star=\bm\Phi^{+}\big(k^\star-\bar k\big),
  \qquad
  u(\cdot\,;\bm\xi^\star)=\sum_{m=1}^{M}u^{m}_{\bm x}(\cdot)\prod_{j=1}^{S}u^{m}_{j}(\xi^\star_j),
  \label{eq:query}
\end{equation}
where $u^m_{\bm x}=\prod_{i=1}^{d}u^m_{x_i}$ is the spatial part of the $m$-th mode; $\bm\Phi$
contains the scaled modes $\phi_j$ at the quadrature points; and $\bm\Phi^{+}$ is its pseudo-inverse,
which is a projection since the modes are orthogonal. An inference requires $S$ one-dimensional
interpolations per mode and one contraction, and the projected coordinates $\bm\xi^\star$ are also used for the
domain check of Section~\ref{sec:query}.

\section{Experimental setup}
\label{app:setup}
In this section, we define each problem, its input field and its reference solution, and then give
the settings of INO and of the baselines.

\subsection{Problem definitions}
\label{app:problems}
All input fields are Gaussian random fields with the squared-exponential covariance
$C(\bm x,\bm x')=\exp\big(-\|\bm x-\bm x'\|^2/2\ell^2\big)$ on the unit domain, where $\ell$ is the
correlation length. A sample can be written as:
\begin{equation}
    g(\bm x;\bm\zeta)=\sum_{j}\zeta_j\,\phi_j(\bm x),\qquad \zeta_j\sim\mathcal N(0,1)\ \text{i.i.d.},
  \label{eq:grf}
\end{equation}
where $\phi_j$ is the $j$-th KL mode scaled by the square root of its eigenvalue, in decreasing order
of the eigenvalues; and $\zeta_j$ is its coordinate. For an input $\bar k+\sigma g$, the first $S$
coordinates $\zeta_j$ are the $\xi_j$ of Eq.~(\ref{eq:kl}). A test sample retains all modes of this
expansion (\NUM{256} in two dimensions and \NUM{533} in three), whereas INO retains only the first
$S$ (Table~\ref{tab:setup}). All methods are evaluated on the same test samples, \NUM{200} per
problem in two dimensions (\NUM{60} at $513^2$), \NUM{40} in three dimensions and \NUM{100} for the
one-dimensional problems. Every reference solution is computed for the true input field and,
except for the moving source, with the same discretization as INO: C-HiDeNN for diffusion and
Helmholtz, and linear elements for the nonlinear diffusion and diffusion--reaction problems
(Table~\ref{tab:setup}).

\paragraph{Moving heat source.} We consider the one-dimensional transient problem:
\begin{equation}
\begin{aligned}
  \partial_t u(x,t)-\partial_{xx}u(x,t)&=P(t)\,\exp\!\big(-2(x-t)^2/\epsilon^2\big),\quad (x,t)\in(0,1)\times(0,1],\\
  u(0,t)&=u(1,t)=0,\qquad u(x,0)=0,
\end{aligned}
  \label{eq:pde-move}
\end{equation}
where $\epsilon=0.01$ is the width of a heat spot that crosses the domain at unit speed. The input is
the power history $P(t)=500+\sigma g(t)$ with $\sigma=150/\sqrt3\approx86.6$ and $\ell=0.1$. INO
retains the \NUM{18} leading KL coordinates, and its parametric factors are the Hermite polynomials
$\mathrm{He}_a(\xi_j)/\sqrt{a!}$, $a\le2$, which are orthonormal under the standard normal density
and contain the exact parametric factors $1$ and $\xi_j$, since the source is affine in $\bm\xi$.
Therefore, the operator has no bounded parametric domain. The reference solution is a
Crank--Nicolson finite difference solution on a finer $2001\times2001$ grid, and errors are relative $\ell^2$ errors on
$500\times51$ points in $(x,t)$.

\paragraph{Diffusion.} We consider steady diffusion with a random coefficient:
\begin{equation}
  -\nabla\cdot\big(k(\bm x;\bm\zeta)\,\nabla u(\bm x)\big)=1,\quad \bm x\in(0,1)^d,\qquad
  u(\bm x)=0,\quad \bm x\in\partial(0,1)^d,
  \label{eq:pde-diffusion}
\end{equation}
with $d=2$ and $3$, where $k=\bar k+\sigma g$ with $\bar k=1$, $\sigma=0.2$ and $\ell=0.3$. The
operator maps $k$ to $u$. The reference solution is computed by the preconditioned conjugate gradient
method, with the operator for the mean input field as preconditioner, until the relative
residual is below $10^{-11}$ ($10^{-10}$ in three dimensions); at $129^2$, it agrees with a sparse
direct solve to \NUM{$5.2\times10^{-12}$}.

\paragraph{Helmholtz.} We consider the Helmholtz equation with a random squared wavenumber:
\begin{equation}
  -\Delta u(\bm x)-\kappa^2(\bm x;\bm\zeta)\,u(\bm x)=1,\quad \bm x\in(0,1)^d,\qquad
  u(\bm x)=0,\quad \bm x\in\partial(0,1)^d,
  \label{eq:pde-helmholtz}
\end{equation}
with $d=2$ and $3$, where $\kappa^2=\bar\kappa^2+\sigma g$ with $\bar\kappa^2=10$, $\sigma=1$ and
$\ell=0.3$. The parametric domain is chosen such that $\kappa^2$ stays below the first Dirichlet
eigenvalue of the Laplacian, $\mu_1=2\pi^2\approx19.7$ in two dimensions and $3\pi^2\approx29.6$ in
three, so that the problem is coercive for every input within the parametric domain. The operator
maps $\kappa^2$ to $u$, and the reference solution is computed as for diffusion.

\paragraph{Nonlinear diffusion.} We consider the 2D nonlinear equation:
\begin{equation}
  -\nabla\cdot\big[(1+\beta\,|\nabla u|^2)\,\nabla u(\bm x)\big]=f(\bm x;\bm\zeta),\quad
  \bm x\in(0,1)^2,\qquad u(\bm x)=0,\quad \bm x\in\partial(0,1)^2,
  \label{eq:pde-quasilinear}
\end{equation}
with $\beta=1$, where $f=\bar f+\sigma g$ with $\bar f=1$, $\sigma=1$ and $\ell=0.3$, so that the
source changes sign within the domain. The operator maps $f$ to $u$. The reference solution is
computed by Newton's method, with each step solved by a sparse direct solver, until the norm of the
residual is below $10^{-10}$.

\paragraph{Diffusion--reaction.} Following \citet{wang2021learning}, we consider the space--time
problem:
\begin{equation}
\begin{aligned}
  \partial_t u(x,t)&=0.01\,\partial_{xx}u(x,t)+0.01\,u(x,t)^2+f(x),\quad (x,t)\in(0,1)\times(0,1],\\
  u(x,0)&=0,\qquad u(0,t)=u(1,t)=0,
\end{aligned}
  \label{eq:pde-dr}
\end{equation}
where the source $f$ is a zero-mean Gaussian random field with $\ell=0.2$, as in
\citet{wang2021learning}. The operator maps $f$ to $u(x,t)$. The reference solution is computed by
Newton's method as for nonlinear diffusion, until the norm of the residual is below $10^{-12}$.

\subsection{Discretization and settings}
For the coefficient problems, INO uses C-HiDeNN with $s=3$, $a=20$, $p=2$ and three quadrature
points per element in every direction (Appendix~\ref{app:chidenn}); for the nonlinear diffusion and
diffusion--reaction problems, it uses linear elements. INO is trained on one CPU core by greedy
enrichment, and the parametric factors of each new mode are initialized with Gaussian bumps whose
centers and widths differ between directions. For the moving source, whose input enters only the
load, all $M=500$ modes are instead updated together by block alternating least squares with five
sweeps. For the nonlinear diffusion problem, the coefficient $1+\beta|\nabla u|^2$ is evaluated for
the previous iterate and frozen. Since this coefficient is itself in separated form, every iteration
is a linear problem of the form of Eq.~(\ref{eq:terms}) with a number of terms independent of $S$,
and the coefficient is compressed to rank \NUM{48} before each iteration. The settings of each
experiment are listed in Table~\ref{tab:setup}.

\begin{table}[h]
\caption{Settings of each experiment. $L$ is the width of the parametric domain in standard
deviations, followed by the number of elements per parametric direction; ``Hermite'' marks Hermite
parametric factors, which have no bounded parametric domain; ``greedy'' marks a rank set by the
greedy enrichment; ``candidates'' is the number of random initial guesses per mode; and
$\alpha_{\Omega}$ is the smallest coercivity constant over the parametric domain.}
\label{tab:setup}
\begin{center}
\footnotesize\setlength{\tabcolsep}{3pt}
\resizebox{\linewidth}{!}{\begin{tabular}{llllll}
\toprule
experiment & mesh & $S$ & domain $L$ / elements & rank $M$ & other \\
\midrule
Table~\ref{tab:neural}, moving source & $2001\times51$ & 18 & Hermite, no bounded parametric domain & 500 & block ALS, 5 sweeps \\
Table~\ref{tab:neural}, Fig.~\ref{fig:bound}, 2D diffusion, Helmholtz & $65^2$--$513^2$ & 25 & 4 / 80 & 96 & $\sigma=0.2$ / $1$, $\bar\kappa^2=10$ \\
Table~\ref{tab:neural}, 3D diffusion & $65^3$, $129^3$ & 56, 164 & 4 / 80 & 300, 160 & 3 candidates per mode \\
Table~\ref{tab:neural}, 3D Helmholtz & $65^3$, $129^3$ & 164 & 4 / 80 & 160 & 3 candidates per mode; $\sigma=1$ \\
Table~\ref{tab:neural}, nonlinear diffusion & $65^2$, $129^2$ (linear elements) & 49 & Hermite, no bounded parametric domain & 295, 297 & coefficient rank 48 \\
Table~\ref{tab:neural}, diffusion--reaction & $65\times65$ & 16 & Hermite & greedy & linear elements \\
Table~\ref{tab:accuracy}, rank study & $65^2$ ($513^2$) & 9 & 2 / 40 & 4--96 & $\sigma=0.2$ \\
Table~\ref{tab:dsweep} & $33^d$ & 8 & 2 / 40 & 64 & $\alpha_{\Omega}=0.45$ \\
\bottomrule
\end{tabular}}
\end{center}
\end{table}

\subsection{Baselines}
\label{app:baselines}
The baselines use the settings of their original papers and are trained on one NVIDIA RTX A6000 GPU
each. PI-FNO follows \citet{li2021physics}, with four Fourier layers of width 64 and 20 Fourier modes
in two dimensions (\NUM{26.2}\,M parameters) and 12 modes in three dimensions (\NUM{226.5}\,M
parameters). On the diffusion, Helmholtz and nonlinear diffusion problems, its output is multiplied by
the mollifier $\sin(\pi x)\sin(\pi y)$ of \citet{li2021physics}, with one sine factor per direction in
three dimensions, so that its predictions vanish on the boundary. It is trained for \NUM{15{,}001} Adam steps with a batch size of 20 and a learning rate
of $10^{-3}$, halved at steps 5k, 7.5k and 10k. PI-DeepONet follows \citet{wang2021learning}, with
branch and trunk networks of five 50-unit layers. It is trained for \NUM{120{,}000} steps with
batches of \NUM{10{,}000} points (\NUM{4{,}000} in three dimensions) and a learning rate decayed by
0.9 every \NUM{2{,}000} steps. The diffusion--reaction benchmark uses the architecture and budget of
\citet{wang2021learning}. The moving-source problem has no published setup, so PI-DeepONet uses the
diffusion--reaction settings, and PI-FNO uses the Burgers configuration of \citet{li2021physics} on a
$513\times513$ grid that resolves the heat spot; both are trained on samples of the same Gaussian random
field as the test samples. PI-FNO receives the input field on the grid on which it is trained, and PI-DeepONet reads it at
$65^2$ sensors in two dimensions and $17^3$ in three dimensions. To compute
the errors, PI-DeepONet is evaluated directly at the quadrature points, and the grid output of
PI-FNO is interpolated to them with the same C-HiDeNN basis as INO.

PI-DeepONet is trained for a fixed number of steps, and each step evaluates the PDE residual at
randomly sampled points instead of on the mesh. Therefore, its training time is short and does not
depend on the mesh. However, on the moving-source, Helmholtz and three-dimensional problems, its
error is larger than that of the mean-input solution. Since its error changes by less than \NUM{10\%}
over the last five evaluations, the training budget is not the cause. The larger error comes
from two features of its original setup in \citet{wang2021learning}. First, the boundary conditions are imposed only by a
penalty term, so they are not satisfied exactly; on two-dimensional diffusion, for example, the
error on the boundary is about \NUM{12\%} of the solution maximum. Second, the residual is evaluated
at only \NUM{100} random interior points per input function, which are sparse in a
three-dimensional domain and rarely fall on the narrow heat spot of the moving source (\NUM{3.7\%}
of the points).

\section{Additional experiments}
\label{app:extra}
\subsection{Rank and mesh resolution}
\label{app:rankmesh}
In this section, we study how the rank error of INO depends on the number of modes $M$ and on the mesh (Table~\ref{tab:accuracy}). For two-dimensional diffusion at
$65^2$ with $S=9$, the rank error decreases from \NUM{$2.9\times10^{-2}$} at $M=4$ to
\NUM{$4.0\times10^{-4}$} at $M=96$. At $M=96$, the rank error is between \NUM{$4.0\times10^{-4}$}
and \NUM{$7.8\times10^{-4}$} on the meshes from $65^2$ to $513^2$, so it hardly depends on the mesh.
In two dimensions, Helmholtz needs fewer modes than diffusion, since its input enters only the $\kappa^2u$ term and
not the Laplacian. In three dimensions, the operators at $129^3$ with $S=8$ and $M=128$ are trained
in \NUM{70\,s} (diffusion) and \NUM{40\,s} (Helmholtz) on one CPU core, and each stores
\NUM{91{,}520} numbers. This is \NUM{23} times fewer than the $129^3=\NUM{2{,}146{,}689}$ nodal values
of a single solution on the same mesh, although the operator gives the solution for every input.

\begin{table}[t]
\caption{Rank error with respect to a reference solution of the same discretization, averaged over
test samples within the parametric domain. Diffusion uses $\sigma=0.2$ and a parametric domain of
$\pm1$ standard deviation, and Helmholtz uses $\bar\kappa^2=10$. Repeated training runs of the same
operator differ by up to \NUM{1.4}$\times$, since the initial guesses of the modes are random.}
\label{tab:accuracy}
\begin{center}
\begin{tabular}{llrrrl}
\toprule
problem & mesh & $S$ & rank $M$ & unknowns & rank error \\
\midrule
2D diffusion & $65^2$  & 9 &  96 &  47{,}904 & \NUM{$4.035\times10^{-4}$} \\
2D diffusion & $513^2$ & 9 &  96 & 133{,}920 & \NUM{$4.599\times10^{-4}$} \\
2D Helmholtz & $65^2$  & 9 &  64 &  31{,}936 & \NUM{$4.98\times10^{-5}$} \\
2D Helmholtz & $513^2$ & 9 &  96 & 133{,}920 & \NUM{$4.97\times10^{-6}$} \\
3D diffusion & $129^3$ & 8 & 128 &  91{,}520 & \NUM{$3.916\times10^{-4}$} \\
3D Helmholtz & $129^3$ & 8 & 128 &  91{,}520 & \NUM{$3.736\times10^{-4}$} \\
\bottomrule
\end{tabular}
\end{center}
\end{table}

\subsection{Evaluation on every test sample}
\label{app:accepted}
Every model is evaluated on every test sample, including the samples outside the
parametric domain. For these samples, INO clips the coordinates to the parametric domain; for the
nonlinear diffusion problem, which has no bounded parametric domain, the samples that fail the
domain check are evaluated without change. Including these samples increases the error of INO by
\NUM{1.1}--\NUM{7.5}$\times$, whereas the error of PI-FNO increases by at most \NUM{18\%}
(Table~\ref{tab:alldraw}).

The domain check uses only the leading coordinates, since the KL modes are scaled by the square roots
of their eigenvalues, which decrease rapidly. For the two-dimensional problems ($\ell=0.3$), the
amplitude of the seventh mode is \NUM{0.29} times that of the first, and that of the 25th mode is
\NUM{0.007} times, so the first six of the $S=25$ modes carry \NUM{88\%} of the variance. Therefore,
clipping a trailing coordinate hardly changes the input field, whereas a check of all 25 coordinates
would keep only $0.9545^{25}\approx31\%$ of the samples.

\begin{table}[t]
\caption{Relative errors on every test sample, including the samples outside the parametric domain.
``mean input'' is the error of the mean-input solution. The cells are defined as in
Table~\ref{tab:neural}; \textbf{bold} marks the most accurate
model.}
\label{tab:alldraw}
\centering
\small\setlength{\tabcolsep}{5pt}
\begin{tabular}{lllll}
\toprule
problem (mesh) & mean input & \INO & PI-FNO & PI-DeepONet \\
\midrule
2D diffusion $65^2$         & \NUM{1.32e-1} & \NUM{1.07e-2 (1.01)} & \NUM{$\bm{6.06}$e-3 (1.43)} & \NUM{5.74e-2 (1.05)} \\
2D diffusion $129^2$        & \NUM{1.32e-1} & \NUM{$\bm{1.05}$e-2 (1.03)} & \NUM{5.48e-2 (1.16)} & \NUM{6.19e-2 (1.13)} \\
2D diffusion $513^2$        & \NUM{1.32e-1} & \NUM{$\bm{1.05}$e-2 (1.02)} & \NUM{1.74e-1 (1.01)} & \NUM{5.61e-2 (1.12)} \\
2D Helmholtz $65^2$         & \NUM{7.07e-2} & \NUM{$\bm{2.96}$e-3 (1.01)} & \NUM{7.29e-3 (1.33)} & \NUM{1.19e-1 (1.47)} \\
2D Helmholtz $129^2$        & \NUM{7.07e-2} & \NUM{$\bm{2.96}$e-3 (1.01)} & \NUM{7.52e-2 (1.60)} & \NUM{1.32e-1 (1.27)} \\
2D Helmholtz $513^2$        & \NUM{6.65e-2} & \NUM{$\bm{2.80}$e-3 (1.02)} & \NUM{9.07e-2 (1.00)} & \NUM{1.20e-1 (1.12)} \\
3D diffusion $65^3$         & \NUM{1.13e-1} & \NUM{1.19e-2 (1.02)} & \NUM{$\bm{2.14}$e-3 (1.48)} & \NUM{2.12e-1 (1.06)} \\
3D diffusion $129^3$        & \NUM{1.14e-1} & \NUM{1.41e-2 (1.07)} & \NUM{$\bm{3.54}$e-3 (3.28)} & \NUM{2.05e-1 (1.06)} \\
3D Helmholtz $65^3$         & \NUM{2.99e-2} & \NUM{1.81e-3 (1.02)} & \NUM{$\bm{1.23}$e-3 (1.78)} & \NUM{4.16e-1 (1.31)} \\
3D Helmholtz $129^3$        & \NUM{2.80e-2} & \NUM{1.76e-3 (1.02)} & \NUM{$\bm{8.22}$e-4 (1.53)} & \NUM{4.20e-1 (1.25)} \\
2D nonlinear diffusion $65^2$  & \NUM{9.60e-1} & \NUM{$\bm{3.08}$e-3 (1.08)} & \NUM{9.13e-3 (2.10)} & \NUM{2.18e-1 (1.06)} \\
2D nonlinear diffusion $129^2$ & \NUM{8.91e-1} & \NUM{$\bm{2.84}$e-3 (1.08)} & \NUM{5.56e-2 (4.11)} & \NUM{2.19e-1 (1.15)} \\
\bottomrule
\end{tabular}
\end{table}

\subsection{Training PI-FNO on fine meshes}
\label{app:finemesh}
\begin{figure}[t]
\begin{center}
\includegraphics[width=0.98\linewidth]{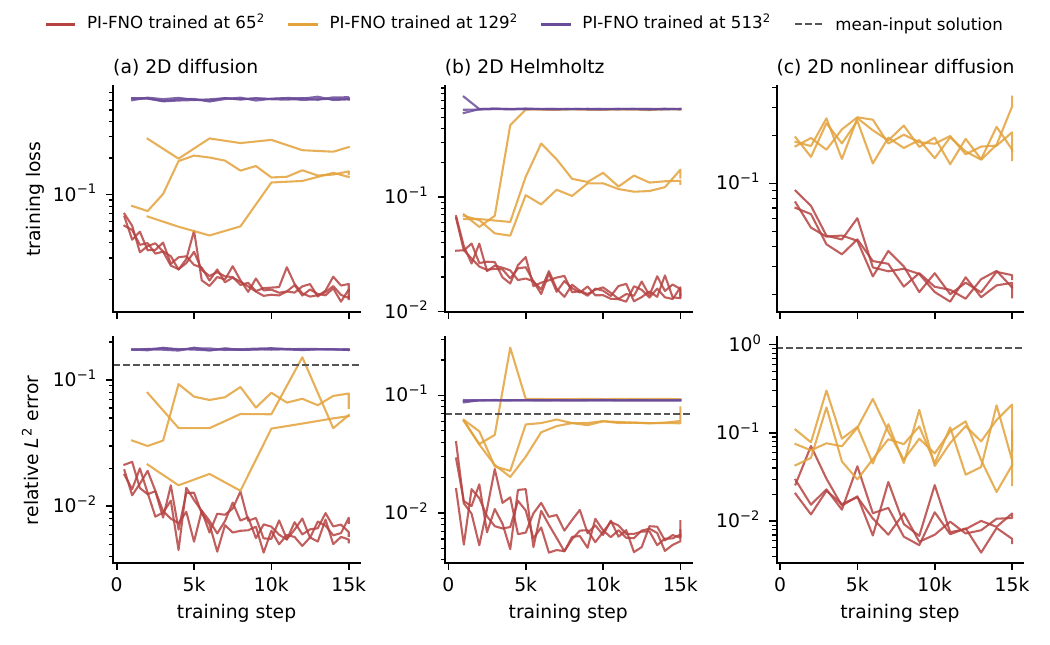}
\end{center}
\caption{Training histories of PI-FNO on the grid of each problem, for the three initializations of
Table~\ref{tab:neural}. Top: training loss. Bottom: relative $L^2$ error on all test samples. The
dashed line is the error of the mean-input solution.}
\label{fig:pifno-training}
\end{figure}

As shown in Fig.~\ref{fig:pifno-training}, with its published settings, the
training loss of PI-FNO decreases steadily at $65^2$. At $129^2$, it increases after a few thousand
steps on diffusion and Helmholtz and hardly decreases on nonlinear diffusion. At $513^2$, neither the
loss nor the error changes after the first \NUM{2{,}000} steps, and the error stays above that of the
mean-input solution. For the same initialization, a smaller learning rate of $3\times10^{-4}$ reduces the diffusion error
at $129^2$ from \NUM{$6.0\times10^{-2}$} to \NUM{$2.6\times10^{-2}$}, which is still larger than at
$65^2$. Since
the Fourier neural operator is resolution invariant, the network trained at $65^2$ can also be
applied on the finer grid. As shown in Table~\ref{tab:finemesh}, it is then more accurate than the
network trained on the fine mesh, and on two-dimensional diffusion it is also more accurate than
INO. This use of the network is not included in Tables~\ref{tab:neural} and~\ref{tab:alldraw}, where
every model is trained on the mesh of the problem for fair comparison.

\begin{table}[t]
\caption{Relative errors of PI-FNO on every test sample when it is trained on the fine mesh and
when it is trained at $65^2$ and applied on the fine mesh (range over three initializations),
compared with INO (geometric mean over three initializations, as in Table~\ref{tab:alldraw}).}
\label{tab:finemesh}
\begin{center}
\small
\begin{tabular}{lllll}
\toprule
problem & mesh & trained at mesh & trained at $65^2$, applied & \INO \\
\midrule
2D diffusion & $129^2$ & \NUM{5.18--6.02e-2} & \NUM{2.53--3.98e-3} & \NUM{1.05e-2} \\
2D diffusion & $513^2$ & \NUM{1.73--1.75e-1} & \NUM{3.14--3.74e-3} & \NUM{1.05e-2} \\
2D Helmholtz & $129^2$ & \NUM{5.81--9.29e-2} & \NUM{5.13--6.09e-3} & \NUM{2.96e-3} \\
2D Helmholtz & $513^2$ & \NUM{9.06--9.07e-2} & \NUM{4.06--4.69e-3} & \NUM{2.80e-3} \\
2D nonlinear diffusion & $129^2$ & \NUM{2.57e-2--1.06e-1} & \NUM{4.02e-3--1.01e-2} & \NUM{2.84e-3} \\
\bottomrule
\end{tabular}
\end{center}
\end{table}

\subsection{Baseline variability}
\label{app:variability}
Across three initializations, the error of PI-FNO changes by \NUM{1.0}--\NUM{4.5}$\times$ and that
of PI-DeepONet by \NUM{1.0}--\NUM{1.9}$\times$, whereas that of INO changes by at most
\NUM{1.4}$\times$ (Tables~\ref{tab:neural} and~\ref{tab:alldraw}). The error of PI-FNO also changes
by up to \NUM{8}$\times$ over its last five evaluations, so its final checkpoint is noisy. In a separate study of the nonlinear diffusion problem with a stronger nonlinearity
($\beta=10$), three PI-FNO initializations differ by a factor of \NUM{10.9}, and the best of them is
more accurate than INO. Therefore, we report every result together with its spread.

\subsection{Inference cost}
\label{sec:cost}
Table~\ref{tab:query-cost} compares the inference cost of INO with that of a conventional numerical
solve. The error bound of Eq.~(\ref{eq:cert}) additionally requires one matrix--vector product with $A(k)$ and one evaluation of the dual norm,
which is inexpensive since $K_0$ is diagonal in the eigenbasis of its one-dimensional factors. For a
true input field, $A(k)$ is not in separated form, so the field is first written in separated form,
and this step dominates the cost of the bound.

\begin{table}[t]
\caption{Inference cost of INO and of a sparse five-point finite difference solve for
two-dimensional diffusion ($S=9$, $M=96$, single-threaded medians). The rows with a bound use
$M=64$ and inputs given by their KL coordinates.}
\label{tab:query-cost}
\centering
\footnotesize
\begin{tabular}{lrrrr}
\toprule
 & $65^2$ & $129^2$ & $257^2$ & $513^2$ \\
\midrule
degrees of freedom          & 4{,}225 & 16{,}641 & 66{,}049 & 263{,}169 \\
inference & \NUM{$23\,\mu$s} & \NUM{$54\,\mu$s} & \NUM{$169\,\mu$s} & \NUM{$623\,\mu$s} \\
finite-difference solve     & \NUM{11.1\,ms} & \NUM{53.3\,ms} & \NUM{281.6\,ms} & \NUM{1.80\,s} \\
ratio                       & \NUM{$484\times$} & \NUM{$991\times$} & \NUM{$1{,}662\times$} & \NUM{$2{,}892\times$} \\
\midrule
inference $+$ rank-error bound & \NUM{0.54\,ms} & \NUM{1.61\,ms} & --- & --- \\
inference $+$ total-error bound & \NUM{5.26\,ms} & \NUM{12.5\,ms} & --- & --- \\
\bottomrule
\end{tabular}
\end{table}

\subsection{Solving for the coefficients}
\label{app:readoff}
\label{sec:readoff}
In Eq.~(\ref{eq:ansatz}), the coefficient of each mode is the product of its parametric
factors at the input coordinates, and the error of INO comes from these coefficients
(Section~\ref{sec:efficiency}). Since the weak
form is a polynomial in $\bm\xi$ and in $u$, the coefficients can instead be obtained for a given
input by solving the weak form projected onto the spatial modes, which is a system of size $r\le M$,
linear for diffusion and cubic for the nonlinear diffusion problem. The projected matrices are
computed once after training, so INO remains data-free. The solved coefficients reduce the error by
\NUM{21.5}$\times$ at $M=96$ and \NUM{124}$\times$ at $M=320$ on two-dimensional diffusion, and by
\NUM{5.3}$\times$ on the nonlinear diffusion problem, at an inference cost of \NUM{150\,$\mu$s}
and \NUM{0.5\,s}, respectively (Table~\ref{tab:readoff}).

\begin{table}[t]
\caption{Errors of the same operator evaluated with the product form of Eq.~(\ref{eq:ansatz}) and
with solved coefficients. The errors are rank errors for diffusion and total errors for the
nonlinear diffusion problem. For diffusion, $S=9$ and the errors are averaged over \NUM{30} samples,
which differ from the samples of the text.}
\label{tab:readoff}
\begin{center}
\small
\resizebox{\linewidth}{!}{\begin{tabular}{lllll}
\toprule
problem & product form & solved coefficients & gain & inference time, product / solved \\
\midrule
2D diffusion $65^2$, $M{=}96$  & \NUM{$3.21\times10^{-4}$} & \NUM{$1.49\times10^{-5}$} & \NUM{21.5$\times$} & \NUM{27\,$\mu$s} / \NUM{150\,$\mu$s} \\
2D diffusion $65^2$, $M{=}320$ & \NUM{$9.45\times10^{-5}$} & \NUM{$7.63\times10^{-7}$} & \NUM{124$\times$} & \NUM{62\,$\mu$s} / \NUM{1.08\,ms} \\
nonlinear diffusion $65^2$             & \NUM{$3.08\times10^{-3}$} & \NUM{$5.79\times10^{-4}$} & \NUM{5.3$\times$}  & \NUM{146\,$\mu$s} / \NUM{0.5\,s} \\
\bottomrule
\end{tabular}}
\end{center}
\end{table}

\subsection{Scaling with spatial dimension}
\label{sec:dsweep}
In this example, we study how the training cost of INO grows with the spatial dimension $d$. The
parametric problem is fixed, and the spatial dimension is varied from $d=1$ to $d=5$ with 33 nodes
per direction. The training time increases only from \NUM{4.2\,s} at $d=2$ to \NUM{7.0\,s} at
$d=5$ (Table~\ref{tab:dsweep}), while the spatial
grid grows by more than four orders of magnitude, since each spatial dimension only adds one
one-dimensional solve per sweep over the directions. The ratio of the bound to the true error stays between \NUM{1.16}
and \NUM{1.42}. At $d=5$, the spatial grid has $3.9\times10^{7}$ points and no reference solution was
computed, but the bound gives a relative energy-norm error of at most \NUM{$3.03\times10^{-3}$}.

\begin{table}[t]
\caption{Cost versus spatial dimension with $n=33$ nodes per direction and $M=64$, with $S=8$ for
$d\ge2$ and $S=4$ for $d=1$, and $\sigma$ chosen such that $\alpha_{\Omega}=0.45$. Five test
samples within the parametric domain are used, and all timings are on one CPU core. Errors and
bounds are relative energy-norm errors, and ``ratio'' is the mean ratio of the bound to the error.
$^{\dagger}$No reference solution was computed at $d=5$.}
\label{tab:dsweep}
\begin{center}
\resizebox{\linewidth}{!}{\begin{tabular}{rrrrrlll}
\toprule
$d$ & spatial DOF $n^d$ & unknowns & terms $T$ & training time (s) & error & bound & ratio \\
\midrule
 1 &            33 & 12{,}608 &  5 & \NUM{2.0} & \NUM{$1.09\times10^{-4}$} & \NUM{$1.27\times10^{-4}$} & \NUM{1.16} \\
 2 &         1{,}089 & 25{,}216 & 18 & \NUM{4.2} & \NUM{$1.71\times10^{-3}$} & \NUM{$2.15\times10^{-3}$} & \NUM{1.25} \\
 3 &        35{,}937 & 27{,}328 & 27 & \NUM{5.5} & \NUM{$5.21\times10^{-3}$} & \NUM{$6.35\times10^{-3}$} & \NUM{1.21} \\
 4 &     1{,}185{,}921 & 29{,}440 & 36 & \NUM{6.0} & \NUM{$2.24\times10^{-3}$} & \NUM{$3.19\times10^{-3}$} & \NUM{1.42} \\
\textbf{5} & 39{,}135{,}393 & 31{,}552 & 45 & \NUM{7.0} & n/a$^{\dagger}$ & \NUM{$3.03\times10^{-3}$} & --- \\
\bottomrule
\end{tabular}}
\end{center}
\end{table}

\end{document}